\documentclass{an}
\usepackage[dvips]{graphicx}
\usepackage{times}
\usepackage{fancyhdr}
\usepackage{epsf}

\begin{document}
\title{The distribution of current helicity at the solar surface
  at the beginning of the solar cycle}
  \author{D.\, Sokoloff\inst{1,2} \and S.D.\, Bao\inst{1} \and
  N.\, Kleeorin\inst{3}
  \and K. \, Kuzanyan\inst{4,5} \and  D.\, Moss \inst{6}
  \and I.~Rogachevskii\inst{3} \and  D.\,~Tomin\inst{7} \and H.\,
  Zhang\inst{1}}
  \institute{National Astronomical Observatories, Chinese Academy of
  Sciences, Beijing 100012, China
  \and Department of Physics, Moscow State University,
  Moscow 119992, Russia
  \and Department of Mechanical Engineering, Ben-Gurion
  University of Negev, POB 653, 84105 Beer-Sheva, Israel
  \and IZMIRAN, Troitsk, Moscow Region 142190, Russia
  \and School of Mathematics, University of Leeds, Leeds LS2 9JT, UK
  \and School of Mathematics, University of Manchester,
  Manchester M13 9PL, UK
  \and Department of Mechanics and Mathematics, Moscow State University,
  Moscow 119992, Russia
}
\date{Received; accepted; published online}
\abstract{A fraction of solar active regions are observed to have
current helicity of a sign that contradicts the polarity law for
magnetic helicity; this law corresponds to the well-known Hale
polarity law for sunspots. A significant excess of active regions
with the "wrong" sign of helicity is seen to occur just at the
beginning of the cycle. We compare these observations with
predictions from a dynamo model based on principles of helicity
conservation, discussed by Zhang et al. (2006). This model seems
capable of explaining only a fraction of the regions with the
wrong sign of the helicity. We attribute the remaining excess to
additional current helicity production from the twisting of rising
magnetic flux tubes, as suggested by Choudhuri et al. (2004). We
estimate the relative contributions of this effect and that
connected with the model based on magnetic helicity conservation.
}
\correspondence{sokoloff@dds.srcc.msu.su} \maketitle

\section{Introduction}

According to the current consensus, the solar cycle is associated
with the propagation somewhere in the solar convective shell of a
wave of magnetic field, the "dynamo wave". The origin of this wave
is dynamo action generated by the solar differential rotation and
the helicity of turbulent convective flows, which drives the
"$\alpha$-effect" introduced by Steenbeck, Krause and R\"adler in
1966 (see Krause and R\"adler, 1980). This concept has been
intensively discussed in the literature for about 50 years,
beginning with the seminal paper of Parker (1955a), and many
important results have been obtained. Until recently however, the
concept has remained to some extent speculative because no direct
observations or laboratory confirmation of the key ingredient of
the process, i.e. the $\alpha$-effect, were available. In the last
12 years or so, observations of current helicity in solar active
regions (Seehafer, 1990; Pevtsov et al., 1994; Longcope et al.,
1998; Zhang and Bao, 1998, 1999) have presented a possibility of
confronting theoretical ideas concerning the $\alpha$-effect with
observational evidence.

The point is that the $\alpha$-effect consists of two
contributions (Pouquet et al., 1976),

\begin{equation}
\alpha = \alpha^v+ \alpha^m, \label{frisch}
\end{equation}
where $\alpha^v$ is determined by the mirror asymmetry of
turbulence and is proportional to the hydrodynamic helicity
$\chi^v = \langle {\bf v} \,\cdot\, {\rm curl\,} {\bf v} \rangle$,
while $\alpha^m$ is determined by the mirror asymmetry of the
turbulent magnetic field and is proportional to the current
helicity density $\chi^c = \langle {\bf j \cdot b} \rangle$. Here
$\bf v$ is the turbulent convective velocity, $\bf b$ is the
small-scale magnetic field and ${\bf j} = {\rm curl\,}{\bf b}$ is
the corresponding electric current. $\langle \ldots \rangle$
denotes averaging over an ensemble of convective pulsations. If
the turbulent convection is considered as locally homogeneous and
isotropic, $\chi^c$ is proportional to the magnetic helicity
density $\chi^m = \langle {\bf a \cdot b} \rangle$, where $\bf a$
is the fluctuation of a magnetic vector-potential. Magnetic
helicity is a non-diffusive integral of motion and a topological
invariant proportional to the linkage number of magnetic field
lines.

Magnetic helicity is bounded from above by the magnetic energy
(Moffatt 1978) and the capacity of the small-scale part of the
magnetic spectrum is too small to allow an effective spectral
transport of magnetic helicity. According to the conventional
scenario, the solar dynamo begins from a state with a weak
magnetic field with correspondingly small magnetic helicity.
Because the large-scale magnetic field participating in the dynamo
wave is helical, its magnetic helicity has to be compensated by
the magnetic helicity (of opposite sign) of a small-scale magnetic
field, which also contributes to $\alpha^m$. Correspondingly,
magnetic helicity conservation effectively constrains the dynamo
action.

On the other hand, $\chi^m$ can be determined from solar
observations because the Zeeman effect as exploited
observationally gives in principle three magnetic field
components. In contrast, the Doppler-effect used for velocity
observations give the line-of-sight velocity only, and no
realistic way to determine $\chi^v$ from observations is known.

Indeed, observations of $\chi^c$ in solar active regions provide
the only direct observational (or experimental) information
concerning the $\alpha$-effect available at the moment. Note that
a non-zero $\alpha$-effect means that the electric current
averaged over convective motions has a component parallel to the
averaged magnetic field while the electric current in conventional
electrodynamics is orthogonal to the magnetic field. This peculiar
property of convection (or turbulence) in rotating electrically
conductive flows obviously requires some observational or
experimental confirmation.

Because the magnetic helicity data provide unique information
concerning the key ingredient of the dynamo, making a comparison
with predictions of dynamo theory looks an attractive proposition.
Such a comparison has been performed by Kleeorin et al. (2003) and
Zhang et al. (2006, hereafter Paper I) and shows that the data
demonstrate something similar to the theoretical predictions. The
discussions presented in these papers stress that the quality of
both the data available and the theoretical models, as well as the
length of the time series, are all rather limited and many obvious
questions concerning the comparison remain obscure.

In particular, current helicity is observed at the solar surface
while the dynamo action occurs somewhere inside the Sun. A
magnetic tube rising  to the solar surface to produce an active
region can be twisted by the Coriolis force and so obtain a
component of current helicity in addition to that generated in the
solar interior. It means that the current helicity data exploited
for comparison with dynamo theory could be biased by another
contribution produced during the rise of the tube to the solar
surface. Of course, the twist of magnetic tubes is interesting by
itself in context of the theory of sunspots.

Note that the Coriolis force does not affect directly the magnetic
and current helicities (i.e. the Coriolis force does not enter the
equation for the evolution of the magnetic and current
helicities). On the other hand, the Coriolis force creates the
kinetic or hydrodynamic helicity in inhomogeneous turbulence, and
the kinetic or hydrodynamic helicity enters the equation for the
evolution of the magnetic and current helicities.

We stress that apart from the magnetic helicity conservation
constraint in the solar dynamo, other possibilities for the
production of current helicity production at the solar surface
have been discussed (see e.g. Bao et al. 2002). In particular,
Longscope et al. (1988) associated the current helicity with the
twisting  of a flux tube during the rise of the tube to the solar
surface. A possible way to estimate the contribution to the
current helicity connected with the tube rise was suggested by
Choudhuri et al. (2004a). They considered the current helicity
production during tube migration and predicted that this
additional current helicity should dominate just at the beginning
of the cycle. According to the theoretical predictions as well as
the observational data, this contribution to the helicity follows
a version of the Hale polarity law, i.e. for the major part of the
active region, the sign of current helicity in the northern solar
hemisphere is opposite to that in the southern hemisphere.  We
stress that the polarity law predicts the behaviour of an average
of the data, while substantial current helicity fluctuations are
expected, which are important from the viewpoint of observations,
theory and direct numerical simulations (this last conclusion is
based on the work of Brandenburg \& Sokoloff, 2002). Choudhuri et
al. (2004a) suggest that one effect of tube migration is to
provide a substantial admixture of active regions which violate
the polarity law just at the beginning of the cycle.

Note that, Choudhuri et al. (2004a) defines a measure of helicity
by using a measure of the twist in the magnetic field lines
$\alpha = ({\rm curl \, } {\bf B})_z / B_z$, see their Eq.~(1) and
corresponding explanation in the text of that paper. This
definition differs from the standard one and further clarification
of this aspect of the model is desirable.

The aim of this paper is to compare the ideas presented in
Choudhuri et al. (2004a) with the observational data  for current
helicity obtained at the Huairou Solar Observing station of the
National Astronomical Observatories of China. We show that the
available data is sufficient to demonstrate a contribution of the
rise of flux tubes to the observed current helicity. According to
our estimates about 20\% of active regions at the beginning of the
cycle have the "wrong" sign of current helicity due to this flux
tube effect. On the other hand, the effect is rather moderate and
localized in time, so that overall the current helicity data
retain their role as a valuable source of information about the
properties of current helicity in the domain of field generation.

\section{Current helicity data for the beginning of the cycle}

The first attempt to isolate the contribution from the rise of
flux tubes from the current helicity data now available was
undertaken in Paper I (see also Choudhuri et al., 2004b). Paper I
concluded however that the current helicity observations studied
in that paper do not allow the isolation of the effect of flux
tube rise because the initial stage of the cycle was not covered
by the observations available at that time.

The observational data used in our analysis were obtained at the
Huairou Solar Observing station of the National Astronomical
Observatories of China. The magnetograph using the FeI 5324 \AA \,
spectral line determines the magnetic field values at the level of
the photosphere. The data are obtained using a CCD camera with
$512 \times 512$ pixels over the whole magnetogram. The entire
image size is comparable with the size of an active region, which
at about $2\times 10^{8}$\,m is comparable with the depth of the
solar convective zone.

The observations are restricted to active regions on the solar
surface and we obtain information concerning the surface magnetic
field and helicity only. Monitoring of solar active regions while
they are passing near the central meridian of the solar disc
enables observers to determine the full surface magnetic field
vector. The observed magnetic field is subjected to further
analysis to determine the value ${\bf \nabla} \times {\vec{ b}}$.
Because it is calculated from the surface magnetic field
distribution, the only electric current component that can be
calculated is $({\bf \nabla} \times {\vec{ b}})_z$.  As a
consequence of these restrictions, the derived observable quantity
is

\begin{equation}
H_c = \langle b_z ({\bf \nabla} \times {\vec{ b}})_z \rangle \; ,
\label{observ}
\end{equation}
where $x,y,z$ are local cartesian coordinates connected with a
point on the solar surface, and the $z$-axis is normal to the
surface.

Until now, the largest available systematic dataset on current
helicity has been accumulated during 10 successive years
(1988-1997)  of observations of active regions, consisting of
records of 422 active regions (Bao \& Zhang 1998). It has been
used for theoretical analysis and further data reduction by
Kuzanyan et al. (2000), Zhang et al. (2002), Kleeorin et al.
(2003), Paper~I and Kuzanyan et al. (2006).

The starting point of this paper is that we introduce new
observational data into the discussion. The new data considered
here covers the three years of the beginning of the solar cycle
23, namely 1998-2000. This dataset was discussed earlier by Bao et
al. (2000, 2002), and contains data for 88 active regions.
\footnote{Note that Bao et al. (2000, 2002) were interested in
comparison of various observed quantities in addition to the
current helicity. All the quantities were determined for 64 active
regions only. Because we focus our attention on the current
helicity, we use the whole dataset.} The new data are obtained by
the same technique and processed in much the same way, as the
earlier dataset of Bao and Zhang (1998) covering the ten year
period 1988-1997, see also Zhang and Bao (1998). Thus we feel it
reasonable to merge these two sets of data and henceforth will
consider them as a single continuous dataset for 510 active
regions.

All of the available data is presented in Fig.~1. This Figure
shows the raw data concerning the sign of the helicity presented
as as a butterfly diagram. "+" denotes positive sign of helicity
and dots negative. Two consecutive cycles are shown, and the
distribution of signs more or less agrees with the polarity law.
However a substantial number of active regions with the "wrong"
sign of helicity can also be seen.

\begin{figure}
\hbox to\columnwidth{
 \hss \epsfysize=10cm
\epsfbox{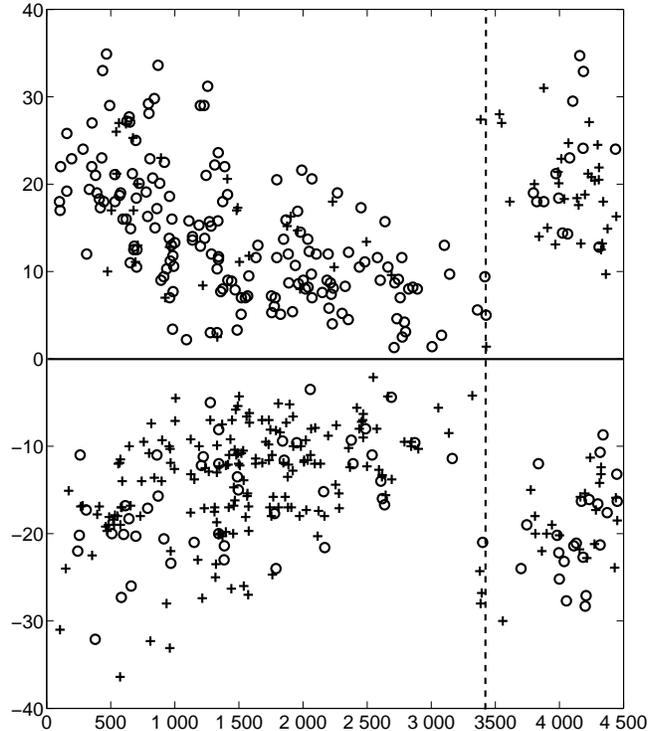}
 \hss}
\caption{Distribution of the sign of magnetic helicity from the
observations at Huairou Solar Station at 1988 - 2000. Time in days
from the beginning of observations is given on the horizontal
axis, and latitude in degrees is given on the vertical axis. Signs
"+" denote an active region with positive current helicity and
circles denote the active region with negative current helicity.
The vertical line separates the old dataset from the new.}
\end{figure}

\section{Active regions with the "wrong" sign of current helicity}

Our aim in the following is to follow the dynamics of the fraction
of the active regions with "wrong" sign of helicity, as identified
in the data presented in Fig.~1. In principle, the problem is
nothing more than a straightforward calculation, comparing the two
types of active region. A few practical points however have to be
fixed.

{\bf The synthetic butterfly diagram.} First of all, note that the
observations cover an interval that is longer  than the cycle
length, and data from two activity cycles are included. The
important point however is that no single cycle is covered
completely and the most interesting part of the cycle, i.e. the
beginning of the cycle, is known from one cycle, while the
behaviour during the main part of the cycle is traced by the
previous cycle. Thus we have to construct a synthetic cycle from
the data.

The procedure used was as follows. We separated the data in the
two cycles by a naked-eye decision. Because the cycle separation
here is sufficiently pronounced we do not feel that anything more
formal is required at the moment. We present the result of this
procedure in Fig.~2.

\begin{figure}
\hbox to\columnwidth{
 \hss \epsfysize=10cm
\epsfbox{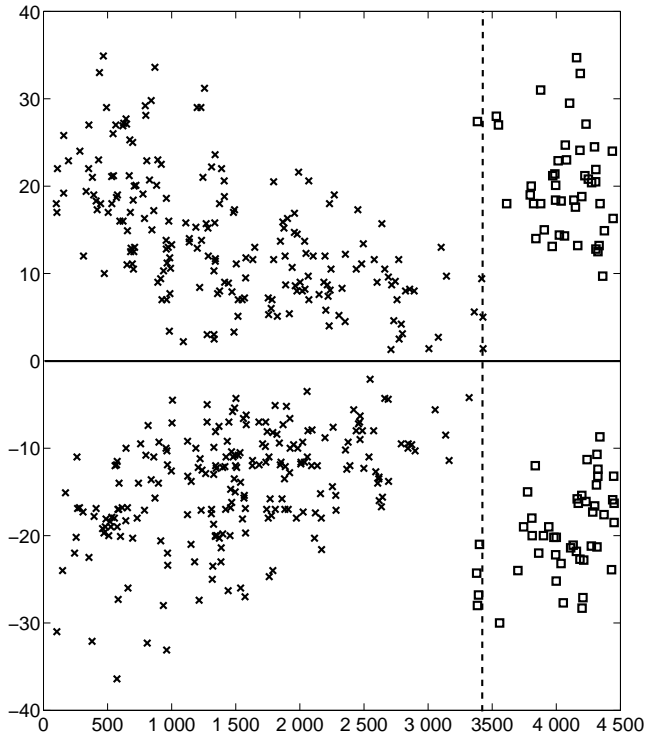}
 \hss}
\caption{Separation of the active regions with known current
helicity over two consecutive cycles. Coordinates are as in
Fig.~1. Crosses denote active regions of the first cycle, while
the boxes indicate active regions from the second cycle.}
\end{figure}

Then we have to shift in time the data from the second cycle to
place them at the beginning of the first cycle. The time-shift $T$
has to be chosen to be equal to the cycle length, which is close
to 11 yr, but is not known precisely {\it a priori}. We tried
several values of $T$ (see Fig.~3) and choose $T= 4000$ d (a value
that is remarkably close to 11 years), based on a naked-eye
estimate of the smoothness of the synthetic butterfly diagram. As
a result, we arrive at the synthetic butterfly diagram shown in
Fig.~4, where the signs of the current helicities are shown
(again, plus signs denote positive helicity and circles the
negative ones).

\begin{figure}
\begin{center}
\includegraphics[width=3.9cm]{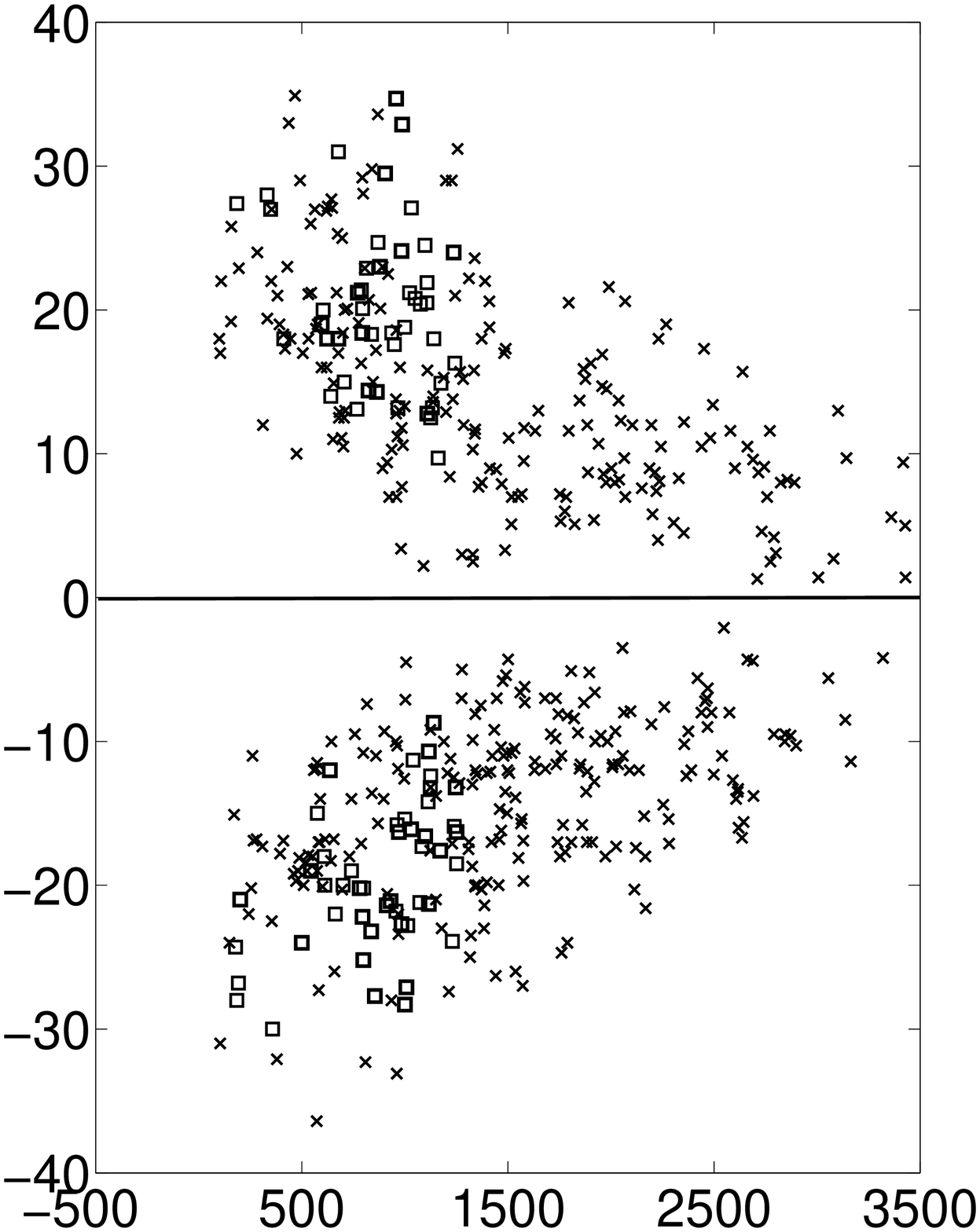}
\includegraphics[width=3.9cm]{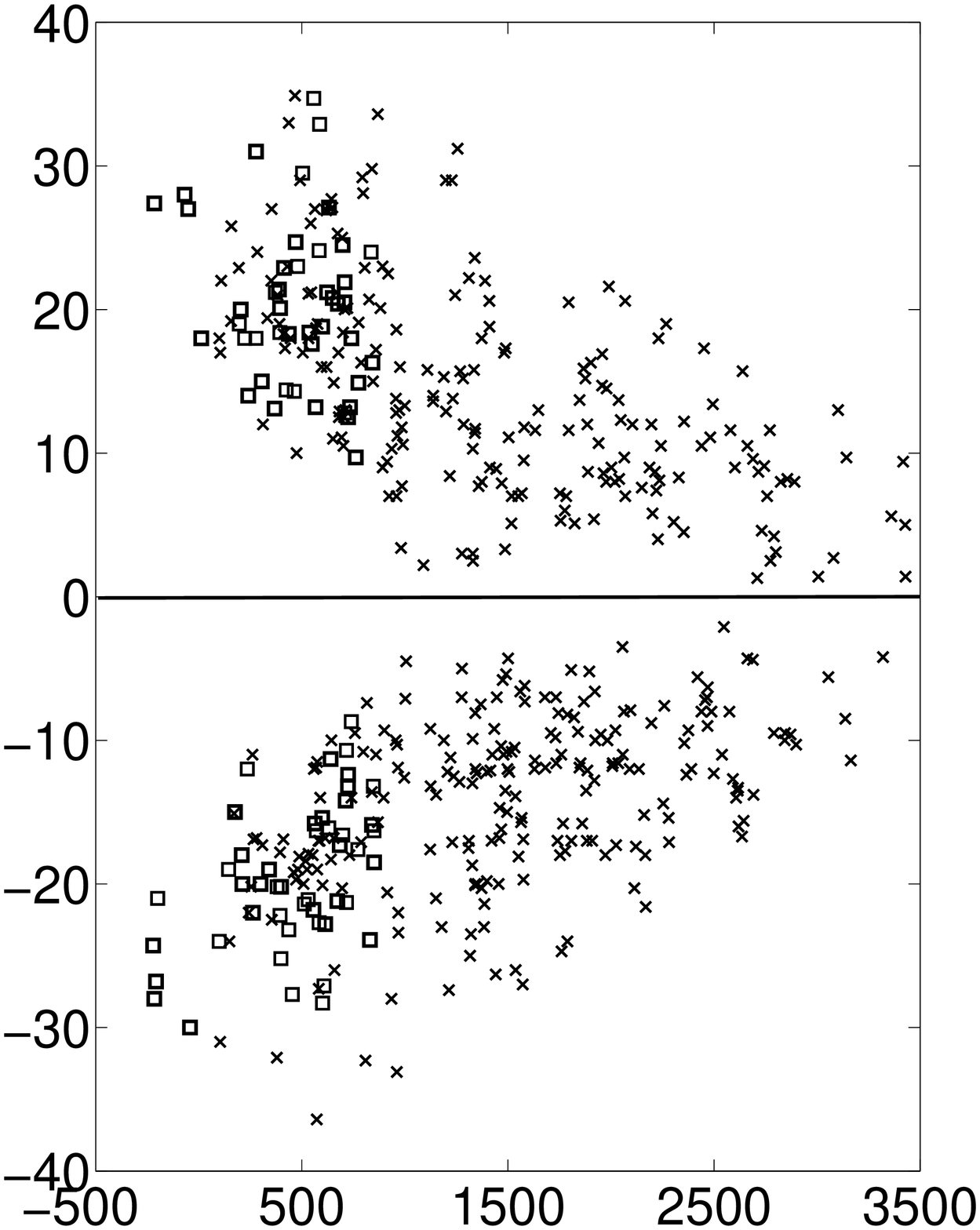}\\
\includegraphics[width=4.2cm]{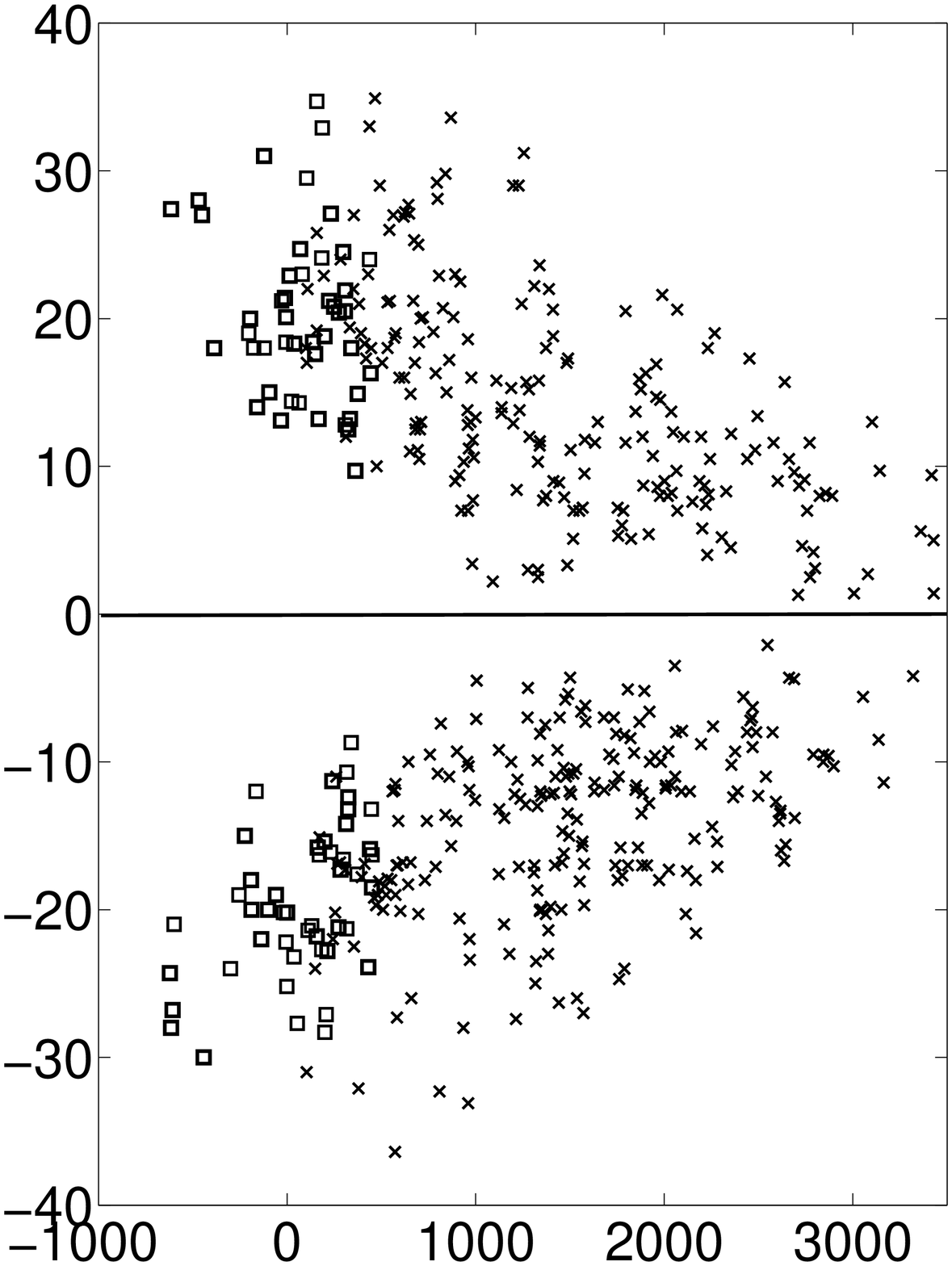
}\\
\includegraphics[width=3.9cm]{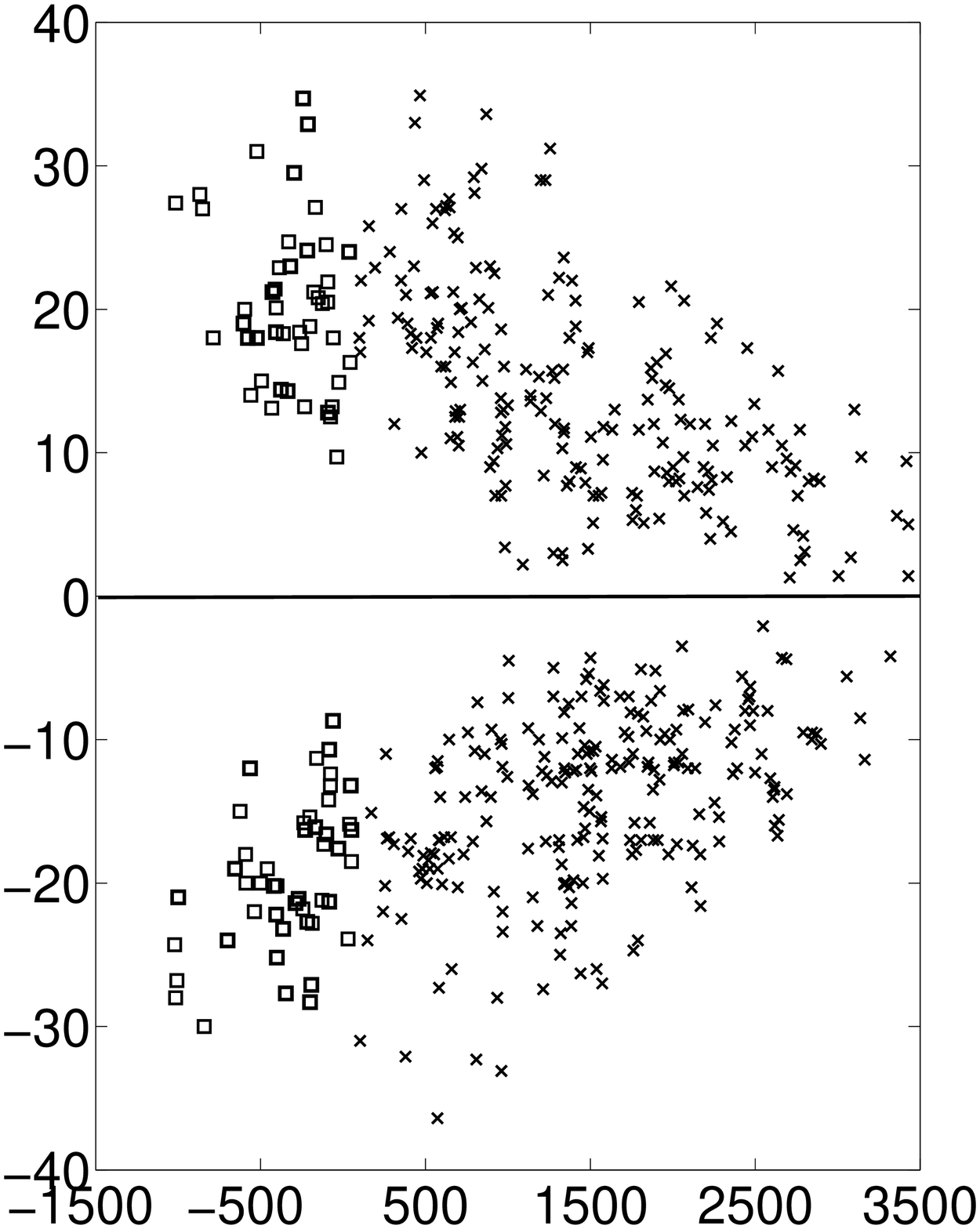}
\includegraphics[width=3.9cm]{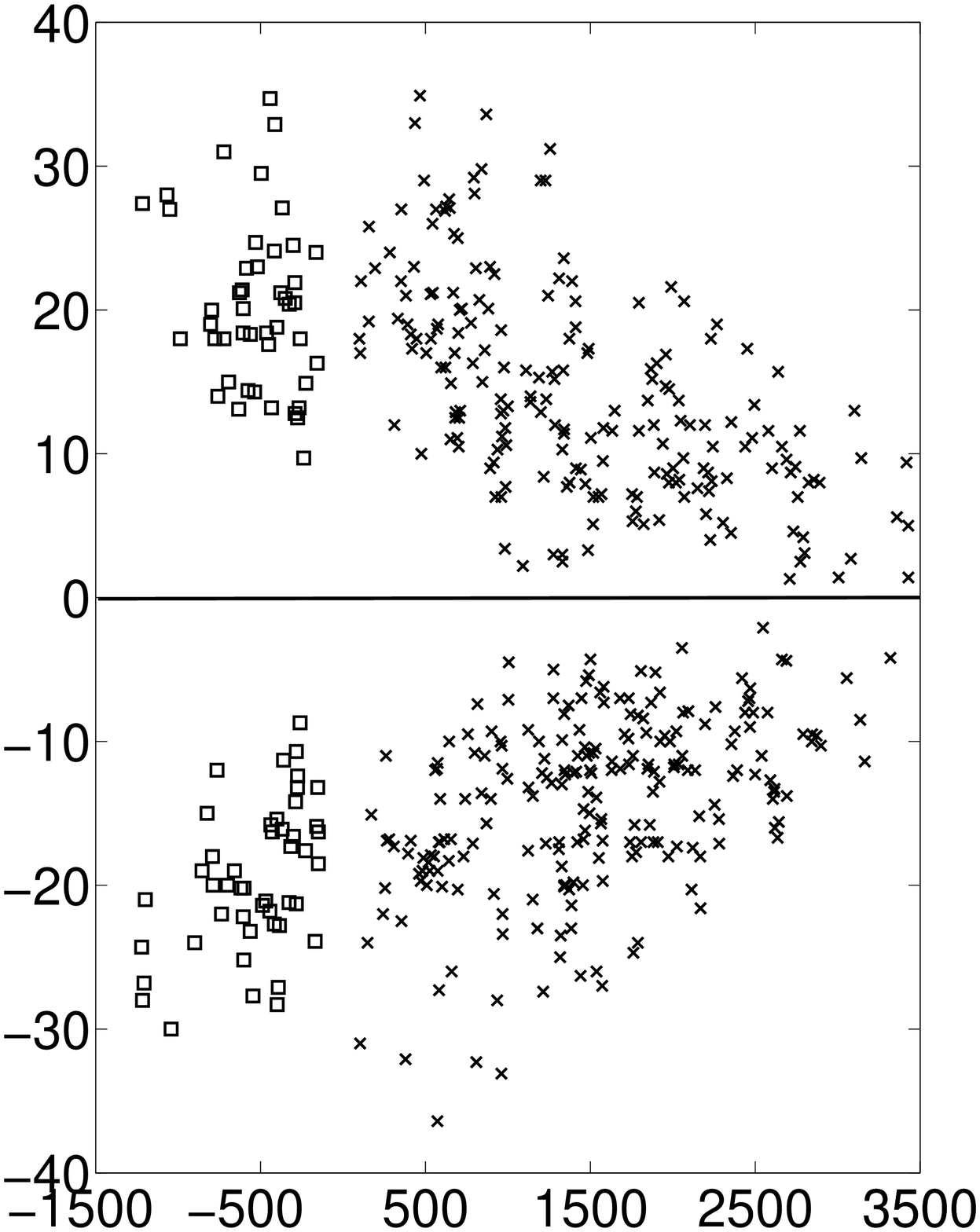}
\caption{Synthetic butterfly diagrams for various time-shifts $T$:
$T= 3200$ days (top row, left);   $T= 3600$ days (top row, right);
$T = 4000$ days {middle row};  $T = 4400$ days (bottom row, left); $T
= 4600$ days (bottom row, right). A time-shift $T= 4000$ days (panel
c) has been chosen as most plausible. Notation is as in Fig.~2.}
\end{center}
\end{figure}

\begin{figure}
\hbox to\columnwidth{
 \hss \epsfysize=10cm
\epsfbox{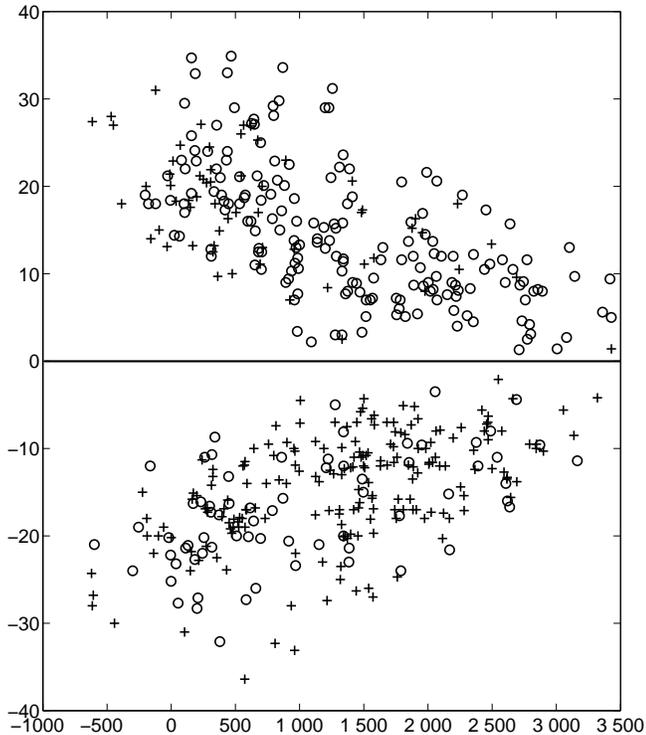}
 \hss}
\caption{Synthetic butterfly diagram with helicity signs. Notation
is as in Fig. 1.}
\end{figure}

{\bf The evolution of the sign of helicity in the synthetic
cycle.} Our aim in the following is to quantify the distribution
shown in Fig.~4. The problem here is as follows. A conventional
procedure would be to divide the temporal extent of the butterfly
diagram into bins and calculate the relative number of regions
with wrong sign (taking into account the polarity law and the
hemisphere in which a given active region is located). The point
however is that the data are quite noisy. If we choose a
reasonable number of bins the number of active regions per bin
drops substantially and the reliability of the results is low.
Thus we use a trick well-known in statistics, but not so familiar
in physics and astronomy. We calculate the cumulative number of
active regions with the "wrong" sign of helicity as well as the
total number of active regions from the beginning of the synthetic
cycle (we are grateful to V.~Tutubalin who suggested this trick to
us). This simple method substantially reduces the noise. We
present the relative number of the active regions with the wrong
sign of helicity in Table~1.

\begin{table}
\begin{tabular}{|l|l|l|l|l|l|l|}

\hline
$t^*$ & $n_-$ & $n_+$ & $p$ & $N_-$ & $N_+$ & $q$\\
\hline
0.18 & 18 & 15 & $54\pm 8 \%$ & 112 & 364 & $24\pm 2 \%$\\
0.30 & 60 & 70 & $46\pm 4 \%$ & 70 & 309 & $18 \pm 2 \%$\\
0.43 & 85 & 144 & $ 37 \pm 3 \%$ & 45 & 235 & $ 16 \pm 2 \%$ \\
0.55 & 101 & 219 & $32 \pm 2 \%$ & 31 & 160 & $15 \pm 3 \%$\\
0.68 & 112 & 289 & $28 \pm 2 \%$ & 18 & 90 & $ 17 \pm 4 \%$\\
0.80 & 121 & 341 & $26 \pm 2 \%$ & 9 & 38 & $19 \pm 5 \%$\\
\hline

\end{tabular}

\caption{Here $t^*$ is the phase of the cycle, i.e. the fractional
time from the beginning of the cycle; $t^* =0$ corresponds to the
beginning of the cycle and $t^* = 1$ to the end. $n_-$ is the
number of active regions with the wrong sign that occur before
phase $T^*$, while $N_-$ means the number of active regions with
the wrong helicity sign occurring after phase $t^*$. The
corresponding notations for the active regions with the "correct"
helicity sign are $n_+$ and $N_+$. The relative numbers of the
active regions before and after phase $T^*$ are $p$ and $q$
respectively. The error bars are calculated as for the Poisson
process.}
\end{table}

We conclude from Table~1 that active regions with the "wrong"
sign of helicity occur preferentially at the beginning of the cycle,
before cycle phase $t^* = 0.175$. Indeed, for $t^* = 0.175$ we
obtain $p=54\%$ while $q \approx 24\%$ for all $t^*$ in the Table.
Note that the data for $t^* = 0.175$ come from the new set of
observations introduced into the analysis in this paper. This is
why we were unable to recognize this phenomenon in the analysis of
Paper~I.

An alternative interpretation of the data in Table~1 would be the
idea that the second cycle included in Table~1 is basically
different from the first with respect to the Hale polarity law for
helicity, and that the new cycle contains more active regions with
the wrong helicity sign than the previous. Although at the moment
we do not see any reason to adopt this interpretation, we stress
that publication of current helicity data from any additional year
of observations would substantially constrain the possible
interpretations. Neither do we see any reason to suggest that the
observational data became much more noisy during the last two
years of observations.

Note that the analysis above of the current helicity data differs
from that undertaken in Paper~I. That paper considered active
regions with known rotation rate, and separated them into deep and
shallow regions according to their rotation rate. A substantial
number of the active regions observed have no reliable depth
identification and were not included in the analysis. As a result,
in Paper~I we were unable to follow the temporal evolution of $p$
and $q$ in detail. Here we do not separate the data by rotation
rate/depth, but add some new data. As a result, we can follow the
evolution of $p$ and $q$, but avoid discussion concerning the
radial distribution of magnetic helicity.

\section{Helicity conservation at the beginning of the cycle}

The natural next step in our analysis is to decide to what extent
the increased percentage of active regions with the "wrong" sign
of helicity at the beginning of the cycle can be instructive for
understanding physical processes within the Sun. We appreciate
that the helicity data currently available are rather crude, and
that any substantial improvement of the data probably lies in the
quite remote future. Correspondingly, we restrict ourself to the
simplest theoretical models (which are really quite non-trivial)
whose complexity is, we feel, more or less comparable with the
state of the data. In particular, we consider the model suggested
by Choudhuri et al. (2004a), alongside  the model developed in
Paper~I, to examine the extent to which the models are compatible
with the behaviour of the active regions with the "wrong" sign of
helicity described above.

We stress that the physical mechanisms underlying these models are
not mutually incompatible. However it is far from  obvious how to
combine them into a synthetic model. The point is that the model
suggested by Choudhuri et al. (2004a) is based on the buoyancy of
the magnetic flux tubes. Magnetic buoyancy applies (in the
astrophysical literature) to two different situations (see Priest
1982). The first corresponds to a problem discussed by Parker
(1966, 1979) and Gilman (1970) who considered a magnetic buoyancy
instability of stratified continuous magnetic field and do not use
the magnetic flux tube concept. The other situation was considered
by Parker (1955b), Spruit (1981), Spruit and van Ballegooijen
(1982), Ferriz-Mas and Sch\"{u}ssler (1993) and Sch\"{u}ssler et
al. (1994). They studied the buoyancy of magnetic flux tubes.
Paper~I included effective velocities which can be considered as
the small-scale magnetic buoyancy of the continuous mean magnetic
field. Therefore, it is not clear at the moment how to combine the
models by Choudhuri et al. (2004a) and Paper I into a synthetic
model. In any case, we feel that such a step would be more than
anything justified by the data now available.

Obviously, the model suggested by Choudhuri et al. (2004a) which
focusses attention on the migration of flux tubes to the solar
surface broadly explains the behaviour under discussion. Note
however that the simulated butterfly diagram for the sign of
current helicity suggested by Choudhuri et al. (2004b) looks
exaggerated, because the active regions with the "wrong" sign are
obviously dominant at the beginning of the cycle. The maximal
corresponding index from Table 1 is $p = 54\%\pm 8\%$ only. The
question is to what extent the model of Choudhuri et al. (2004a)
can explain the other features of the observed helicity
distribution investigated by Kleeorin et al. (2003) and Paper~I.
However, such a study is beyond the scope of this paper.

We use below a two-dimensional axisymmetric nonlinear dynamo model
which includes an explicit radial coordinate, and takes into
account the curvature of the convective shell and density
stratification. The nonlinear model takes into account algebraic
quenching of the total $\alpha$-effect and turbulent magnetic
diffusivity. We split the total $\alpha$-effect into its
hydrodynamic and magnetic parts. The calculation of the magnetic
part of the $\alpha$ effect is based on the idea of magnetic
helicity conservation and the link between current and magnetic
helicities. In the model we use a dynamical equation for magnetic
helicity which includes production, transport (helicity fluxes)
and molecular dissipation of magnetic helicity (see Paper~I for
details).

The model of Paper~I, based on magnetic helicity conservation,
does not appear in principle incompatible with the data (here and
below we use the model suggested by Paper~I without modification).
We demonstrate this by the following simple experiment. We take
two butterfly diagrams, for the deep and shallow domains of the
model of Paper~I, for some particular choice of parameters and
formally combine them with an arbitrary weighting. For example, we
show in Fig.~4 the result from combining a "deep" butterfly
diagram (Fig.~6 of Paper~I, weighted at 0.8), with a surface
diagram (Fig.~7 of Paper~I with weight 0.2). This figure  looks
quite similar to the data from Table~1,
 and as convincing as the plot presented in
Choudhuri et al. (2004b). We stress however that physically the
contributions from the deep and shallow domains cannot be arbitrarily combined
as independent contributions to a butterfly diagram, and a deeper
analysis is required.

\begin{figure}
\includegraphics[width=8cm,height=6cm]{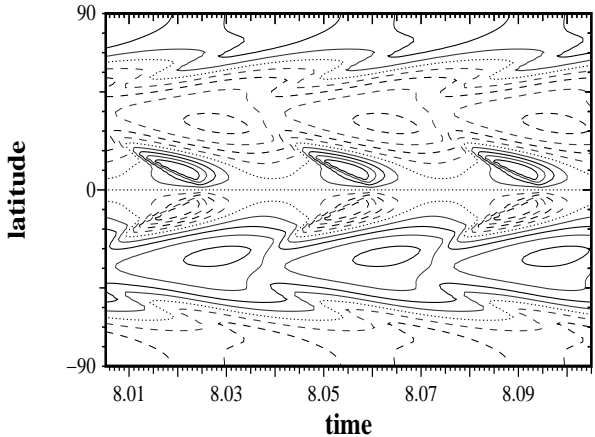}
\caption{Artificial butterfly diagram for the current helicity
obtained by combining diagrams from deep and shallow domains.
Contours of positive values are shown as solid curves, negative
values are broken, and the zero contour is dotted.}
\end{figure}

Analysis of the data obtained from a dynamo model computed as in
Paper~I for a quite typical set of values of the governing
parameters proceeds as follows. (Specifically, the model has
$C_\alpha=-5$, $C_\omega=6\times 10^4$, $T$ varies between 5 and
$5\times 10^4$ from top to bottom of the convection zone, and the
density parameter $a=0.3$. Further details can be found in
Paper~I.) From the solution in the computational box defined by
radial ($r$), co-latitudinal ($\theta$) and time ($t$)
coordinates, we identify a domain associated with a particular
activity wave. We performed this identification based on common
sense arguments and naked-eye estimates. We tried several
prescriptions for this separation of the data. Because the
overlapping of the activity waves in the simulated (as well as
observed) butterfly diagrams is quite modest, the results seem
quite robust with respect to the particular choice of separation
procedure. We omit here presentation of a set of rather similar
tables, but recognize that a more systematic method of separation
of the data would be highly desirable.

Following Paper~I, we identify the the relative number of the
active regions with the "wrong" current helicity sign with the
relative volume of the computational box with the "wrong" helicity
sign. More precisely, we introduce the value $i_- (t^*)$ as a
relative volume of the computational box with the "wrong" sign of
the current helicity, up to cycle phase $t^*$, while $n_-$ is the
relative volume of the computational box with the wrong sign after
phase $t^*$.

Of course, the values $i_-$ (to be compared with $p$ from Table~1)
and $n_-$ (to be compared with $q$ from Table~1) depend on the
governing parameters of the model, but the general shape of the
behaviour seems to be quite robust. More details concerning the
computation of $i_-$ and $n_-$ are given in Paper~I. The maximal
values of $i_-$ are about 20\%. We present typical values of $i_-$
and $n_-$ in Table~2.

\begin{table}
\begin{tabular}{|l|l|l| l| l|}
\hline
& \multicolumn{2}{|l|}{$0.64<r<0.80$} & \multicolumn{2}{|l|}{$0.64<r<1$}\\
\hline
$t^*$& $i_-$& $n_-$ & $i_-$ & $n_-$\\
\hline
0.18 & 20\% & 15\% & 6\% & 5\%\\
0.30 & 17\% & 16\% & 5\% & 5\%\\
0.42 & 14\% & 19\% & 4\% & 6\%\\
0.55 & 13\% & 28\% & 4\% & 10\%\\
0.68 & 14\% & 40\% & 4\% & 15\%\\
0.80 & 15\% & 56\% & 5\% & 21\%\\
\hline
\end{tabular}
\caption{Relative cumulative volumes occupied by the current
helicity with the "wrong" sign: $i_-$ - before the phase $t^*$,
$n_-$ - after the phase $t^*$. The data are given separately for
the lower domain of the computation box ($0.64 < r <0.80$) and the
whole radial extent of the computational box ($0.64 < r < 1$).}
\end{table}

We see from this Table that the behaviour of $i_-$ is quite
different from that of $p$, e.g. $p$ decreases with $t^*$.  We
were able to find governing parameters which gives some decay of
$i_-$ at the beginning of the cycle but $i_-$ then increases at
the end of the cycle. In addition, the typical values of $i_-$ are
substantially lower then those for $p$. Note that $i_-$ becomes
larger in the lower domain of the computational box ($0.64 < r <
0.80$). The values of $i_-$ become much smaller if the whole
computational box is considered ($0.64<r<1$). The results for
$n_-$ are naturally connected with those for $i_-$.

We conclude from this comparison that our model based on the
magnetic helicity conservation cannot by itself reproduce details
of the behaviour of the index $p$ just at the beginning of the
cycle. In the context of our modelling, this behaviour must be
attributed to the additional current helicity produced by the rise
of flux tubes to the solar surface.

\section{Results and discussion}

We conclude from the above analysis that, from comparison with the
model of Paper~I that the current helicity data from solar active
regions are consistent with a clear contribution from the helicity
production during the rise of magnetic flux tubes to the solar
surface in the formation of active regions. Based on the data
available and theoretical modelling we can give an
order-of-magnitude estimate for the various contributions to the
sign of the surface helicity. About 15\% of the cases with
helicity of the "wrong" sign can be attributed to helicity of the
"wrong" sign originating in the generation domain (this figure is
obtained from Table~2 as a typical value for $i_-$). About
20\%-30\% of the cases with the "wrong" helicity sign must be
attributed to the processes associated with the flux tube rise
(estimated as a difference between maximal and minimal values of
$p$) and the remainder, about 10\%, is attributed to observational
noise.

The idea that the twisting of rising magnetic flux tubes leads to
the effect being discussed looks interesting and promising in the
context of the physics of active regions. From the viewpoint of
solar dynamo theory the effect appears as a bias, but its role is
limited to the beginning of the cycle, and is rather modest. The
current helicity data retains its importance as a unique source of
information about the solar $\alpha$-effect. Taking into account
that the domain of field generation is spatially separated from
the region observed, and also other observational problems (see
details in Kleeorin et al. 2003 and Paper~I), the current helicity
data seem to be surprisingly useful for comparison with
theoretical interpretations.

Our analysis in this paper is not directed towards an
investigation of the radial location of the generation domain. Our
results do however support the localization of the domain deep
inside the convective shell (cf. left and right hand columns of
Table~2; the mechanism of Choudhuri et al. 2004a is also
associated with a deep location of the generation domain).

In spite of the obvious role of twisting processes, magnetic
helicity conservation appear to be responsible for a substantial
fraction of the active regions with the "wrong" sign of helicity.
In particular, we note that an increase of $q$ at the very end of
the cycle (Table~1) might be compared with the growth of $i_-$ at
the end of the cycle (Table~2).

Note that the tendency of the sign of current helicity to reverse
at the beginning of solar cycle was mentioned by Hagino and
Sakurai (2005), from current helicity data obtained at the Solar
Flare Telescope at Mitaka and the Solar 65-cm telescope at
Okayama. The time variation of the sign of current helicty was
inferred from the vector magnetograms observed at the solar
surface. They connected this phenomenon with the inherent
properties of the twisted magnetic field originating from the
solar subatmosphere.  An opposing interpretation was suggested by
Pevtsov et al. (2001) who attributed the tendency to an
observational effect caused by Faraday rotation (see however the
analysis of Bao et al., 2000). We appreciate that the problem
needs further clarification and believe that a systematic
comparison of the data obtained by various observational groups
can provide a crucial contribution in the towards this end.

We stress again the preliminary nature of our findings. Our
results are constrained by the limited extent and quality of the
available observational data as well as by the limited
understanding of the role of current helicity in solar activity at
the moment. Whilst recognizing that future progress in theory and
observations may well lead to a revision of our conclusions, we
nevertheless believe that the results above can stimulate progress
in the problem and, in particular, can provide real constraints
for theories of the solar cycle.

\acknowledgements

The research was supported by grants 10233050, 10228307,
10311120115 and 10473016 of the National Natural Science
Foundation of China, and TG 2000078401 of the National Basic
Research Program of China. DS and KK are grateful for support from
the Chinese Academy of Sciences and NSFC towards their visits to
Beijing, as well as RFBR-NNSFC 02-02-39027 and 05-02-39017. KK
would also like to acknowledge support from RFBR,  grants
03-02-16384 and 05-02-16090. DS is grateful for financial support
from INTAS by grant 03-51-5807 and RFBR by grant 04-02-16068.

\end{document}